\documentclass[apj]{emulateapj} \usepackage{apjfonts} \newif\ifdraft \draftfalse \newif\ifsub \subfalse \newif\ifpre \pretrue
\usepackage{graphicx}
\usepackage{epsf}

\ifdraft
\slugcomment{ 2005.10.15\_17:55:02 \quad To appear in Astronomical Journal}
\fi

\shorttitle{The Fourth VLBA Calibrator Survey}
\shortauthors{Petrov et al.}
\received{2005 August 23}
\revised{2005 October 20}
\accepted{2005 October 22}
\journalinfo{Astronomical~journal}

\ifpre
  \submitted{}
\else
  \paperid{AJ204973}
\fi
\bibpunct{(}{)}{,}{a}{}{,}
\makeindex
\citeindextrue

\newcommand{\getlength}[1]{\ifx#1\end \let\next=\relax
            \else\advance\count255 by1 \let\next=\getlength\fi \next}
\newcount\switch
\newcommand{\Endmat}{\ifnum\switch=0$\fi}
\newcommand{\ifnularg}[1]{ \count255=0 \getlength#1\end \ifnum\count255=0 }
\newcommand{\ifm}{\makebox{}\ifmmode}
\newcommand{\Deg}[1]{#1^\circ}
\long\def\ifundefined#1#2#3{\expandafter\ifx\csname
  #1\endcsname\relax#2\else#3\fi}
\newcommand{\beq}   { \begin{eqnarray} }
\newcommand{\eeq}[1]{ \ifnularg{#1} end{eanarray} \else
                      \label{#1}\end{eqnarray}    \fi }
\newcommand{\eref}[1]{(\ref{#1})}
\newcommand{\Frac}[2]{\frac{\displaystyle\strut #1}{\displaystyle\strut #2} }

\newcommand{\nntab}[2]{ \multicolumn{2}{#1}{#2} }
\newcommand{\nnntab}[2]{ \multicolumn{3}{#1}{#2} }

\newcommand{\Hem}{\hphantom{mm}}

\begin{document}

\title{The Fourth VLBA Calibrator Survey --- VCS4}

\author{L. Petrov}
\affil{NVI, Inc., 7257 Hanover Pkw., Suite D, Greenbelt, MD 20770, USA}
\email{Leonid.Petrov@lpetrov.net}
\author{Y. Y. Kovalev\altaffilmark{\dag}}
\affil{National Radio Astronomy Observatory,
       P.O.\ Box 2, Green Bank, WV 24944, USA;}
\affil{Astro Space Center of Lebedev Physical Institute,
       Profsoyuznaya 84/32, 117997 Moscow, Russia}
\email{ykovalev@nrao.edu}
\author{E. B. Fomalont}
\affil{National Radio Astronomy Observatory,
       520 Edgemont Road, Charlottesville, VA~22903--2475, USA}
\email{efomalon@nrao.edu}
\author{D. Gordon}
\affil{Raytheon/NASA GSFC, Code 697, Greenbelt, MD 20771, USA}
\email{dgg@leo.gsfc.nasa.gov}
\altaffiltext{\dag}{Jansky Fellow, National Radio Astronomy Observatory}

\begin{abstract}

  This paper presents the fourth extension to the Very Long Baseline Array
(VLBA) Calibrator Survey, containing 258 new sources not previously
observed with very long baseline interferometry (VLBI). This survey, based on
three 24 hour VLBA observing sessions, fills remaining areas on the sky above
declination $-40\degr\ $ where the calibrator density is less than one source
within a 4\degr\ radius disk at any given direction. The share of
these area was reduced from 4.6\% to 1.9\%. Source positions were derived
from astrometric analysis of group delays determined at 2.3 and 8.6~GHz
frequency bands using the Calc/Solve software package. The VCS4 catalogue of
source positions, plots of correlated flux density versus projected
baseline length, contour plots and fits files of naturally weighted CLEAN
images, as well as calibrated visibility function files are available on the
Web at \url{http://gemini.gsfc.nasa.gov/vcs4}.

\end{abstract}


\section{Introduction}
\label{s:introduction}

   This work is a continuation of the project of surveying the sky for
bright compact radio sources. These sources can be used as phase referencing
calibrators for imaging of weak objects with very long baseline interferometry
(VLBI) and as targets for space navigation, monitoring the Earth's rotation,
differential astrometry and space geodesy. The method of VLBI, first proposed
by \cite{mat65}, allows us to determine positions of sources with nanoradian
precision (1~nrad $\approx$~0.2~mas). Several catalogues were compiled
combining observations under various programs. The catalogue of sources
observed under geodetic programs from 1979 through 2004,
ICRF-Ext2 \citep{icrf-ext2-2004}, contains positions of 776 sources.
In addition to that, positions of 2247 sources were determined in the
framework of the VLBA Calibrator Survey project: VCS1 \citep{vcs1},
VCS2 \citep{vcs2} and  VCS3 \citep{vcs3}. Since 364 sources are listed
in both the ICRF-Ext2 and the VCS catalogues, the total number of sources
for which positions were determined with VLBI is 2659. Among them,
2269 sources, or 85\%, are considered as acceptable calibrators: they had
at least 8 successful observations at both X and S bands and the semi-major
axis of the error ellipse of their coordinates is less than 25~nrad (5~mas).

  However, the sky coverage of these sources is not uniform. Successful phase
referencing requires a calibrator within at least 4\degr\ from a target
source with a precise position and known source structure. The probability
of finding a calibrator from the combined ICRF-Ext2, VCS1, VCS2 and VCS3
catalogues within 4\degr\ of any target above $ -40\degr\ $ is 95.4\%.
In this paper we present an extension of the VCS catalogues, called the VCS4
catalogue, mainly concentrating on the other 4.6\% of the sky where the
source density is the lowest and on the brightest sources with flat spectrum
previously not observed with VLBI under geodesy and astrometry programs.
Since the observations, calibrations, astrometric solutions and imaging are
similar to that of VCS1--3, most of the details are described by \cite{vcs1}
and \cite{vcs3}. In section \S\ref{s:probability} we assess an a~priori
probability of source detection. In section \S\ref{s:selection} we describe
the strategy for selecting 412 candidate sources observed in three 24 hour
sessions with the very long baseline array (VLBA) based on analysis
of probabilities of source detection which is a further development of the
statistical approach for source selection introduced by \citet{vcs3}.
In section \S\ref{s:obs_anal} we briefly outline the observations and
data processing. We present the catalogue in \S\ref{s:catalogue}, and
summarize our results in \S\ref{s:summary}.

\section{Assessment of an a~priori probability of source detection}
\label{s:probability}

   Having unlimited resources one could try to observe all sources stronger
than some limiting flux density. However, only sources with bright compact
components can be detected with VLBI and may be useful for phase referencing or
geodetic applications. \citet{kellermann} first showed that the distribution
of sources over spectral index $\alpha \enskip (F \propto \nu^{+\alpha})$ has
two peaks: one near $\alpha=-1$ (steep spectrum) and another near $\alpha=0$
(flat spectrum). Later, it was confirmed  that extended objects dominate
in the steep spectrum population \citep[see the review by][]{ko88}.
In compact regions, which have the synchrotron mechanism of emission,
the peak in the spectrum caused by synchrotron self-absorption has
frequencies higher than 1~GHz due to their small size \citep{slysh}.
Thus, if the dominating emission comes from compact regions, the spectrum
of the total flux density will be predominantly flat or inverted.

\begin{figure}[t]
   \ifsub
         \epsscale{0.5}\plotone{f1.eps}
         \par\vspace{-1ex}\par
      \else
         \centerline{ \epsfclipon \epsfxsize=80mm \epsffile{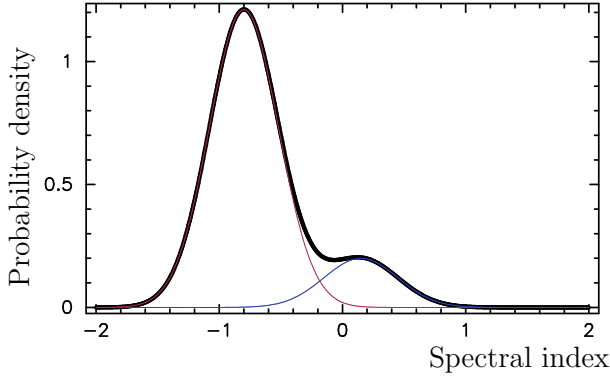} }
   \fi

   \figcaption{ \ifsub \footnotesize \fi \ifdraft \footnotesize \fi
            The probability density distribution at 8.6~GHz as a function
            of spectral index. The thin lines show the distribution of the
            steep spectrum source population (left) and the flat spectrum
            the source population (right).
            \label{f:prob1}
           }
\end{figure}

  Let's determine the probability of detecting with the VLBA a source
with a given total flux density and given spectral index. The empirical
probability distribution function of sources in spectral indexes can be 
approximated as
\beq
    P(\alpha) = (1-a) \, N(\alpha, \alpha_\mathrm{s}, \sigma_\mathrm{s}) \, + \,
                 a \, N(\alpha, \alpha_\mathrm{f}, \sigma_\mathrm{f}) \,,
\eeq{e:e1}
where $a$ is the fraction of the flat-source population to the total
population, $N(\alpha, \alpha_o, \sigma_o )$ is the normal distribution with
maximum at $\alpha_o$ and dispersion $\sigma_o^2$. The first term
$N_\mathrm{s}(\alpha) = N(\alpha, \alpha_\mathrm{s}, \sigma_\mathrm{s})$
represents the probability distribution for steep spectrum sources,
the second one
$N_\mathrm{f}(\alpha) = N(\alpha, \alpha_\mathrm{f}, \sigma_\mathrm{f})$
--- for flat spectrum sources. Using results of \cite{RATAN_NCP} for the
RATAN--600 simultaneous broad-band spectra survey at centimeter wavelengths
of sources around the North Celestial Pole, as well as 2.7 \& 5~GHz data
from \cite{pkscat90a} PKSCat90 catalog with limiting flux density at 2.7
GHz of 250~mJy, we have found the following estimates of distribution
parameters: $a = 0.15$, $\sigma_\mathrm{f} = 0.30$, $\alpha_\mathrm{f} = 0.15$,
$\sigma_\mathrm{s} = 0.28$, $\alpha_\mathrm{s} = 0.80$. Figure~\ref{f:prob1} 
shows the distribution.

  If a source has a spectral index in the range of
$[\alpha, \alpha + d \alpha]$, the ratio of the probability that it belongs
to the flat spectrum population, $P_\mathrm{f}(\alpha)$, to the probability
that it belongs to the steep spectrum population, $P_\mathrm{s}(\alpha)$, is
equal to the ratio of their probability densities:
\beq
  \Frac{P_\mathrm{f}(\alpha)}{P_\mathrm{s}(\alpha)} =
      \Frac{a \, N_\mathrm{f}(\alpha)}{(1-a) \, N_\mathrm{s}(\alpha)} \,.
\eeq{e:e2}
   Since we assume that a source either belongs to the flat spectrum
population or to the steep spectrum population,
$ P_\mathrm{f}(\alpha) + P_\mathrm{s}(\alpha) = 1$.
Then we immediately find
\beq
     P_\mathrm{f}(\alpha) = \Frac{a\, N_\mathrm{f}(\alpha)}
           {(1-a)\, N_\mathrm{s}(\alpha) + a \, N_\mathrm{f}(\alpha)} \,,
     \vspace{2ex} \\
     P_\mathrm{s}(\alpha) = \Frac{(1-a)\, N_\mathrm{s}(\alpha)}
            {(1-a)\, N_\mathrm{s}(\alpha) + a \, N_\mathrm{f}(\alpha)} \,.
\eeq{e:e3}

\begin{figure}[t]
   \ifsub
         \epsscale{0.5}\plotone{f2.eps}
         \par\vspace{-1.5ex}\par
      \else
         \centerline{ \epsfclipon \epsfxsize=78mm \epsffile{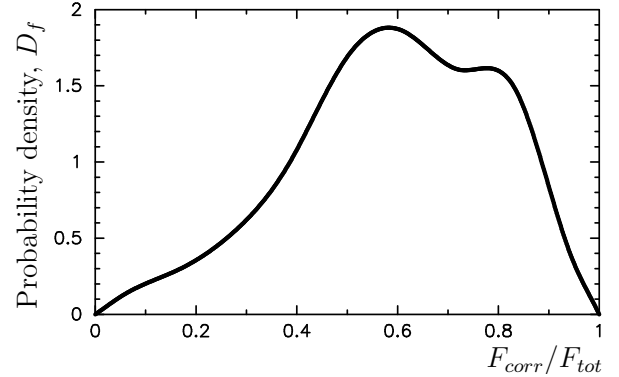} }
   \fi

   \figcaption{ \ifsub \footnotesize \fi \ifdraft \footnotesize \fi
            The probability density distribution at 8.6~GHz of the ratio
            of the correlated flux density at the baseline projections equal
            to the Earth radius to the total flux density among the flat
            spectrum source population.
            \label{f:prob2}
           }
\end{figure}

  The probability that a given source will have a ratio $r$ of the correlated
flux density $F_{corr}$ to the total flux density $F_{tot}$ is quite
different for these two populations. Results of \cite{Kovalev_etal2005} 
of the 2\,cm VLBA survey \citep{2cm} and results of analysis of the VCS3 
observing campaign provided us estimates of $F_{corr}$ at baseline projections 
equal to the Earth radius. The total flux of these sources was measured with 
the RATAN--600. The histogram over 500 sources after smoothing and normalization
gives us an estimate of the empirical probability density distribution 
of the ratio of correlated flux density at baseline projections equal to the 
Earth radius to the total flux density among the flat spectrum source 
population (figure~\ref{f:prob2}).

\begin{figure}[ht]
   \ifsub
         \epsscale{0.5}\plotone{f3.eps}
         \par\vspace{-1.5ex}\par
      \else

         \centerline{ \epsfclipon \epsfxsize=80mm \epsffile{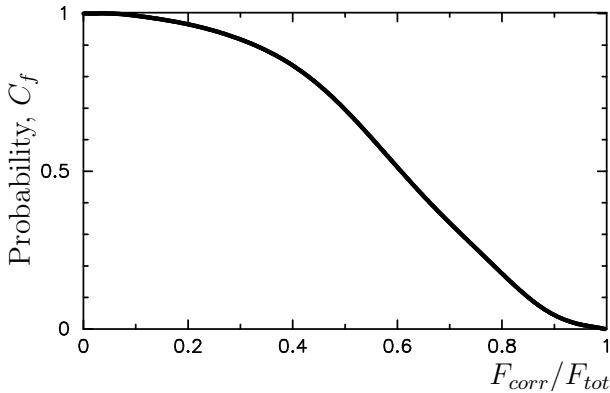} }
         \par\vspace{1.5ex}\par
   \fi

   \figcaption{ \ifsub \footnotesize \fi \ifdraft \footnotesize \fi
            The cumulative probability function of the ratio of the
            correlated flux density to the total flux density among
            the flat spectrum source population:
            $ C_f(r, F_{corr}/F_{tot} ) = P_f(r \le F_{corr}/F_{tot} )$.
            \label{f:prob3}
           }
\end{figure}

  According to our previous experience in processing snapshot VLBA Calibrator
survey observations with integration time 4~minutes and bit rate 128
Msamples/sec with good weather conditions, sources with correlated
flux density greater than 60~mJy are reliably detected. We set the detection
limit of the VLBA to $ F_{lim}$ = 70~mJy  in order to have a little
allowance for possible pointing errors and higher than normal system
temperature due to bad weather. The probability to have the correlated flux
density greater than this limit, $F_{lim}$, i.e. the probability of detection
for a source which belongs to the flat spectrum population and has total
flux density $F_{tot}$ is $ \rho_\mathrm{f}(F_{tot}) =
   C_\mathrm{f} \biggl( \Frac{F_{lim}}{F_{tot}} \biggr) $ where $C_f(r)$ is
the cumulative probability function defined as
\beq
     C_\mathrm{f}(r) = 1 - \int\limits_0^r D_\mathrm{f}(r) \, dr \,,
\eeq{e:e4}
   and where $D_f(r)$ is the probability distribution function of the ratio
of the correlated flux density to the total flux density and $r$ is the
ratio of minimal detected correlated flux density to total flux density.
The plot of $D_\mathrm{f}(r)$ is shown in figure~\ref{f:prob2} and plot
of $C_\mathrm{f}(r)$ is shown in figure~\ref{f:prob3}.

  For steep spectrum sources the probability density function of the ratio
of the correlated flux density to the total flux density is not well known.
According to our analysis of compact steep spectrum sources detected 
in VCS1--3 experiments, we approximate
\beq
   D_\mathrm{s}(r) =
      \cases{
              20,            & \mbox{if} \enskip $ r <    0.05 $\,, \cr
              \hphantom{2}0, & \mbox{if} \enskip $ r \geq 0.05 $\,. \cr
            }
\eeq{e:e5}

  Analogously, we introduce the function $C_\mathrm{s}(r)$ for the steep
spectrum source population:
\beq
     C_\mathrm{s}(r) = 1 - \int\limits_0^r D_\mathrm{s}(r) \, dr \,.
\eeq{e:e6}

  Then, assuming the spectral index of a source is precisely known,
the probability to be detected at 8.6~GHz for a source with spectral
index $\alpha$ and total flux density $F_o$ measured at frequency $f_o$ is
\beq
    P(F_o,f_o,\alpha) =
     P_\mathrm{f}(\alpha) \,
         C_\mathrm{f} \biggl( \frac{F_{lim}}{F_o +  (8.6/f_o)^\alpha} \biggr) +
    \ifpre \nonumber \\ \fi
     P_\mathrm{s}(\alpha) \,
         C_\mathrm{s} \biggl( \frac{F_{lim}}{F_o +  (8.6/f_o)^\alpha} \biggr) \,.
    \ifpre \hphantom{\,.} \fi
\eeq{e:e7}

\begin{figure*}[ht]
   \par\vspace{-3mm}\par
   \ifsub
        \epsscale{1.0}\plotone{f4.eps}
      \else
        \centerline{ \epsfclipon\epsfxsize=\textwidth \epsffile{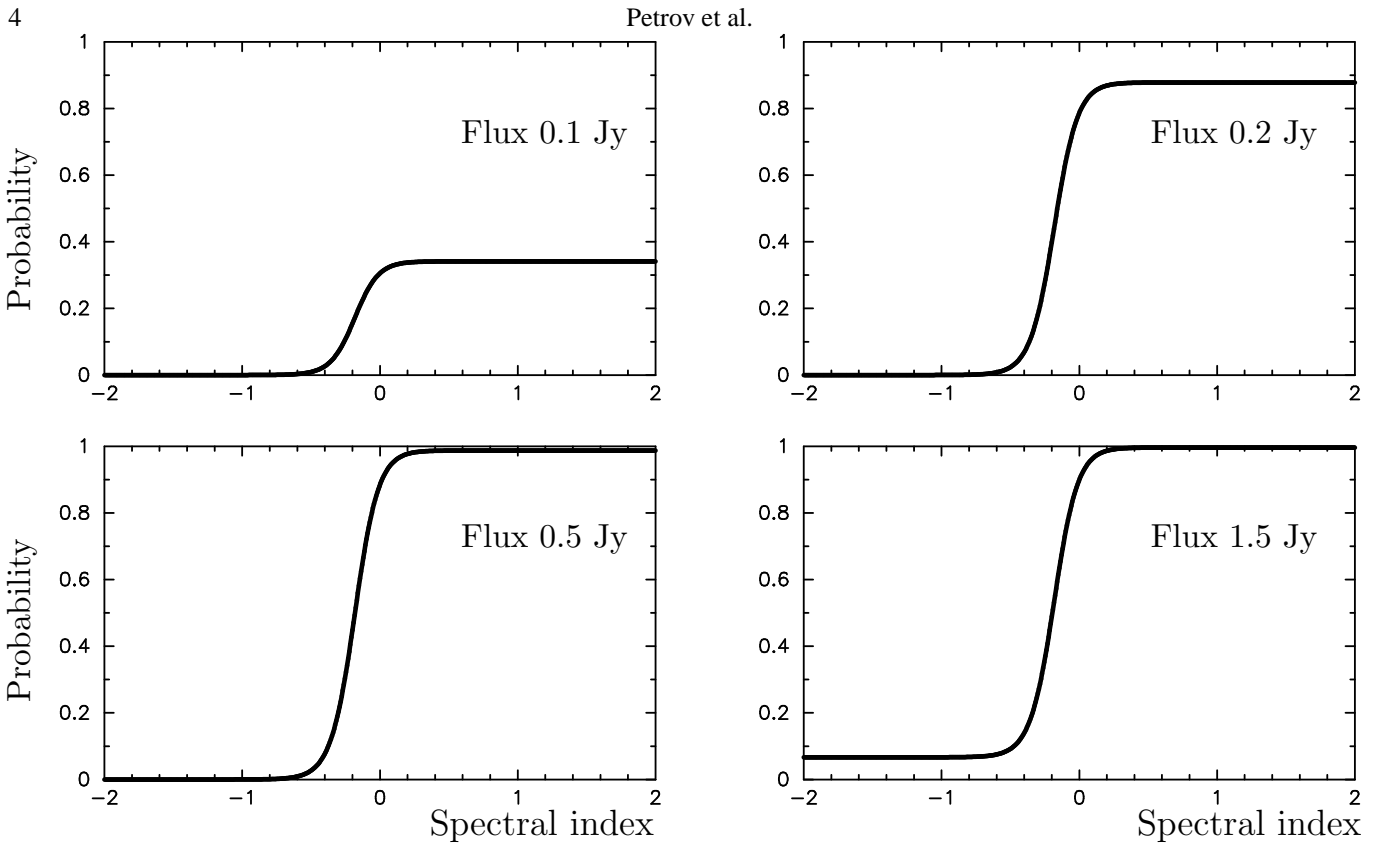}}
   \fi

   \figcaption{ \ifsub \footnotesize \fi \ifdraft \footnotesize \fi
            Probability of being detected at 8.6 GHz as a function of
            spectral index for sources with flux density at 1.4~GHz 0.1~Jy,
            0.2~Jy, 0.5~Jy and 1.5~Jy.
            \label{f:prob4}
           }
\end{figure*}

  Plots of the probability of being detected as a function of spectral
index are presented in figure~\ref{f:prob4}.

\begin{figure}[ht]
   \ifsub
         \epsscale{0.5}\plotone{f5.eps}
         \par\vspace{-1.5ex}\par
      \else
         \centerline{ \epsfclipon \epsfxsize=80mm \epsffile{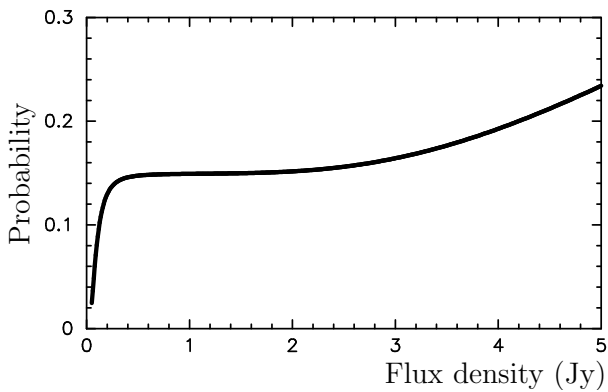} }
   \fi

   \figcaption{ \ifsub \footnotesize \fi \ifdraft \footnotesize \fi
            The probability of detection of a source with unknown
            spectral index as a function of total flux density at 1.4~GHz.
            \label{f:prob5}
           }
\end{figure}

  If no information about the spectral index is available, the estimate
of the probability to detect a source is (see figure~\ref{f:prob5})
\beq
    P(F_o) = \int\limits_{-\infty}^{+\infty}
       \left( P_\mathrm{f}(\alpha) \,
          C_\mathrm{f}\biggl( \frac{F_{lim}}{F_o + (8.6/f_o)^\alpha} \biggr) +
          \ifpre \hphantom{d a \,.} \right. \nonumber \\ \left. \fi
          P_\mathrm{s}(\alpha) \,
          C_\mathrm{s}\biggl( \frac{F_{lim}}{F_o + (8.6/f_o)^\alpha} \biggr)
       \right)
       \, d \alpha \,.
\eeq{e:e8}

\section{Selection of candidate sources for zones with low calibrator
density}
\label{s:selection}

  As the first step of source selection we computed the sky source
density on a $6'\times6'$ grid and identified areas with declination
$ > \Deg{-40} $ with no calibrator within a $\Deg{3}$ disk.

  As the second step we selected all sources from the NVSS catalogue with
flux density greater than 100~mJy at 1.4~GHz, with galactic latitude
$ |b| > \Deg{1}\!.5 $ which either were not previously observed with VLBI
under astrometry and geodesy programs, or which were observed, but did
not collect enough data in order to be counted as calibrators. We used the
CATS database \citep{cats} to gather all available flux density measurements
and then computed a spectral index for each source. Sources with spectral
index less than $-0.5$ were deselected. We also deselected sources with
extrapolated total flux density at 8.6~GHz less than 100~mJy. Sources
for which the spectral index could not be reliably determined were removed
from the list of candidates if they had flux densities at 1.4~GHz less than
200~mJy. We also removed pairs of sources if the distance between components
was less than $\Deg{0}\!.083$, since it usually indicates a source with
extended structure. Those sources, which according to NASA Extragalactic
Database (NED), were listed as galactic objects, H~II regions
or planetary nebulae, were removed as well. Finally, we generated the input
list of 2479 candidates. For 1216 objects we had estimates of their spectral
indexes, and we computed a~priori probability of their detection using
expression \eref{e:e7}. For other sources we computed this probability
using expression \eref{e:e8}. It should be mentioned that many sources have
variable flux density and spectral index. These variations were not taken into
account in evaluation of the probability distribution and equation~\eref{e:e7}.
This makes the estimates of the probability of detection somewhat less
reliable.

  If all these sources could be observed, the mathematical expectation of
the area without a calibrator within a disk with radius of $\Deg{4}$ would
be 0.57\%. For economical reasons we were forced to reduce the list of
candidates. In order to find a subset of $N$ sources in such a manner that
the area with the source calibrator density less than one source in
a disk with radius $\Deg{4}$ would be minimal, we adopted the following
iterative procedure. Initially, we selected all sources. For each point
on the grid with declination $ > \Deg{-40}, $ which {\bf before} the VCS4
campaign had no calibrators closer than $\Deg{4}$, we computed the
probability of that that point will have no calibrator closer than
$\Deg{4}$ { \bf after} VCS4 observations as
\beq
    H_{ij} = \: \prod\limits_{k=1}^{k=n} ( 1 - P_i(\alpha,F) ) \: d(i,j,k) \,,
\eeq{e:e9}
   where $d(i,j,k)$ is 1, if the distance between the point on the grid
$(i,j)$ and the $k$th source is less than $\Deg{4}$, and 0 otherwise.
   The sum
\beq
R = \sum\limits_{grid}^{} H_{ij}
\eeq{e:e10}
  is the mathematical expectation of the total area which has no calibrator
within $\Deg{4}$.

   Then at each step of iteration for each source we compute
$R_k = \displaystyle\sum\limits_{grid}^{}
\mbox{$R \: d(i,j,k)$}\, /\, \mbox{$(1 - P_i(\alpha,F))$}$, i.e., the
contribution of the $k$th source to $R$. The source with the minimal
contribution is removed, the probability $H_{ij}$ is updated, and the
procedure is repeated until $N$ source remained in the list.

  We generated the final list of candidates by concatenating two lists:
1) a list of 300 sources selected according to the iterative process
described above; 2) a list of 100 sources north of declination $\Deg{-30}$
with the highest probability of detection --- these are the strongest
flat spectrum sources previously not observed in geodetic VLBI mode.
We also added 12 sources which had been previously observed and detected
under geodetic programs, but never imaged with the VLBA. Of them, 5 sources
fell in the areas with no calibrators, and therefore were considered as
belonging to list~1, the other 7 sources were added to list~2.
The mathematical expectation of the area which would remain with no calibrator
after observing the final list of 412 objects was 1.77\%.

  The rational for including the list of the one hundred brightest sources
was that the area of the sky without a calibrator within a disk with radius
$\Deg{4}$ is reduced too slowly with an increase in the number of candidates
beyond several hundred. For most geodetic applications having a bright
calibrator a little bit further from the target is more important than having
a calibrator closer, but much weaker. On the other hand, for accurate phase
referencing, a relatively close weak calibrator is often preferable
to a stronger calibrator which is much further away. So, some compromise
is needed in order to improve the calibrator list.

\section{Observations and data processing}
\label{s:obs_anal}

  The VCS4 observations were carried out in three 24 hour observing
sessions with the VLBA on 2005 May 12, 2005 June 12, and 2005 June 30. Each
target source was observed in two scans. The scan duration was 120 seconds for
sources from the list of 107 bright objects, and 235 seconds for other target
sources. The target sources were observed in a sequence designed to minimize
loss of time from antenna slewing. In addition to these objects, 93 strong
sources were taken from the GSFC astrometric and geodetic catalogue
\mbox{\tt 2004f\_astro}\footnote{\url{http://gemini.gsfc.nasa.gov/solutions/astro}}.
Observations of 3--4 strong sources from this list were made every 1--1.5
hours during observing sessions. These observations were scheduled in such
a way, that at each VLBA station at least one of these sources was observed at
an elevation angle less than 20\degr, and at least one at an elevation
angle greater than 50\degr. The purpose of these observations was to provide
calibration for mismodeled atmospheric path delays and to tie the VCS4
source positions to the ICRF catalogue \citep{icrf98}. The list of
troposphere
calibrators\footnote{\url{http://gemini.gsfc.nasa.gov/vcs/tropo\_cal.html}}
was selected among the sources which according to
the 2cm survey results \citep{Kovalev_etal2005} showed the greatest
compactness index, i.e. the ratio of the flux density of an unresolved detail
to the total flux density at VLBA baselines. In total, 505 sources were 
observed. The antennas were on-source 60\% of the time.

  Similar to the previous VLBA Calibrator Survey observing campaign
\citep{vcs3}, we used the VLBA dual-frequency geodetic mode, observing
simultaneously at 2.3~GHz and 8.6~GHz. Each band was separated into four 8~MHz
channels (IFs) which spanned 140 MHz at 2.3~GHz and 490 MHz at 8.6~GHz,
in order to provide precise measurements of group delays for astrometric
processing. Since the a~priori coordinates of candidates were expected
to have errors of up to 30\arcsec, the data were correlated with an
accumulation period of 1~second in 64~frequency channels in order to provide
extra-wide windows for fringe searching.

  Processing of the VLBA correlator output was done in three steps. In the
first step the data were calibrated and fringed using the Astronomical Image
Processing System (AIPS) \citep{aips}. In the second step data were imported
to the Caltech DIFMAP package~\citep{difmap} and images were produced using
the optimized automated procedure originally suggested by Greg
Taylor~\citep{difmap-script}. We were able to reach the VLBA image thermal
noise
level\footnote{\url{http://www.vlba.nrao.edu/astro/obstatus/current/obssum.html}}
for most of our CLEAN images \citep{VLBA_summ}. Errors of our estimates
of correlated flux density of sources brighter than 100~mJy are
determined mainly by the accuracy of amplitude calibration, which for the
VLBA, according to \citet{VLBA_summ}, is at the level of 5\% at 1--10~GHz.
This estimate was confirmed by comparison of the correlated flux density
with the single-dish flux density which we measured with \mbox{RATAN--600}
in March 2005 for slowly varying sources without extended structure. The
methods of single-dish observations and data processing can be found in
\citet{Kovalev_etal99}. In the third step, the data were imported to the
Calc/Solve software, group delays ambiguities were resolved, outliers
eliminated and coordinates of new sources were adjusted using ionosphere-free
combinations of X~band and S~band group delay observables
of the 3 VCS4 sessions, 15 VCS1--3 experiments and 3976 twenty four hour
International VLBI Service for astrometry and geodesy (IVS)
experiments\footnote{\url{http://gemini.gsfc.nasa.gov/solutions/2005c}} in
a single least square solution. The boundary conditions which require zero
net-rotation of new coordinates of the 212 sources listed as defined in the
ICRF catalogue with the respect to their positions from that catalogue were
imposed in order to select a unique solution of differential equations
of photon propagation.

\begin{figure}[ht]
   \ifsub
         \epsscale{0.5}\plotone{f6.eps}
         \par\vspace{-1ex}\par
      \else
         \centerline{ \epsfclipon \epsfxsize=80mm \epsffile{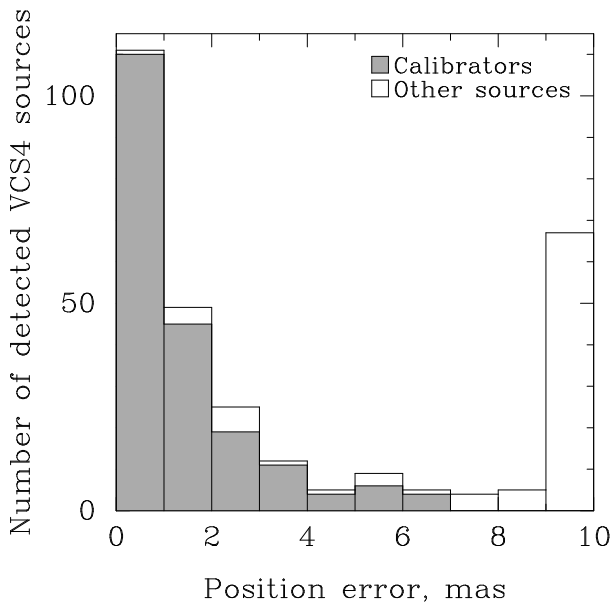} }
   \fi

   \figcaption{ \ifsub \footnotesize \fi \ifdraft \footnotesize \fi
            Histogram of semi-major error ellipse of position errors.
            The last bin shows errors exceeding 9~mas.
            \label{f:hist}
           }
\end{figure}

  In a separate solution, coordinates of the 93 well known tropospheric
calibrators were estimated from the VCS4 observing sessions only. Comparison
of estimates of coordinates of these sources with coordinates derived from
analysis of 3976 twenty four hour IVS experiments provided us a measure of
the accuracy of coordinates from the VCS4 observing campaign. The differences
in coordinate estimates were used for computation of parameters $a$ and
$b(\delta)$ of error inflation model in the form of
$ \sqrt{ (a\, \sigma)^2 + b(\delta)^2}$, where $\sigma$ is an uncertainty
derived from the fringe amplitude signal to noise ratio using the error
propagation law and $\delta$ is declination. More details about analysis and
imaging procedure can be found in \citet{vcs1} and \citet{vcs3}. The histogram
of source position errors is presented in Figure~\ref{f:hist}.

  In total, 292 out of 412 sources were detected and yielded at least
two good points for position determination. However, not all of these sources
are suitable as phase referencing calibrators or as targets for geodetic
observations. Following \citet{vcs3} we consider a source suitable as
a calibrator if 1) the number of good X/S pairs of observations is 8 or greater
in order to rule out the possibility of a group delay ambiguity resolution
error; and 2) position error before reweighting is less than 5~mas following
the strategy adopted in processing VCS observations. Only 196 sources satisfy
this calibrator criteria. It should be mentioned that our criterion for
suitability as a phase calibrator is rather conservative, and sources which
fail this criterion may still be useful for some applications. Among 292
detected sources, 34 were previously observed but did not have enough
observations before the VCS4 campaigns to be considered as calibrators.
Of these 34 objects, new VCS4 observations of 27 of them provided enough
information to classify them as calibrators.

  Table~\ref{t:rate} shows statistics of the a~priori and a~posteriori
detection rate. The a~priori probability gives us a reasonable estimate of the
detection rate. The sources from list~1 are weaker than the sources from
list~2, and for 40\% of them the integration time of 4 minutes was
insufficient to qualify them as calibrators. The a~posteriori detection rate
turned out greater than the a~priori rate, because we set the detection limit
to 70~mJy when computing the a~priori detection rate. In fact, we were able
to detect sources with correlated flux density as low as 50~mJy with
integration time 4~minutes.

\begin{table}[t]
  \caption{ \footnotesize\rm The number of observed and detected sources}
  \begin{center}
  \begin{tabular}{l @{\enskip} r @{\enskip} r @{\enskip} r
                 \ifpre @{\quad} \else @{\qquad} \fi
                  r @{\enskip} r @{\enskip} r }
     \hline
       Data   & \# sources & \# detections & \# calibrators              &
                                             \nntab{c}{detection rate}   &
                                             calibrators \\
              &            &               &                             &
                                             a~priori     &
                                             a~posteriori & \\
     \hline
     List 1   &   305 &   199 &   117  &  45\,\% &  65\,\% &  38\,\% \\
     List 2   &   107 &    93 &    82  &  84\,\% &  87\,\% &  77\,\% \\
     Total    &   412 &   292 &   199  &  57\,\% &  71\,\% &  48\,\% \\
     \hline
  \end{tabular}
  \end{center}
  \label{t:rate}
\end{table}

\ifpre
\begin{figure*}[ht]

   \ifsub
        \epsscale{0.90}\plotone{f7.eps}
      \else
        \centerline{ \epsfclipon
                     \ifdraft \epsfxsize=0.95\textwidth \fi
                     \ifpre \epsfxsize=1.00\textwidth \fi
                     \epsffile{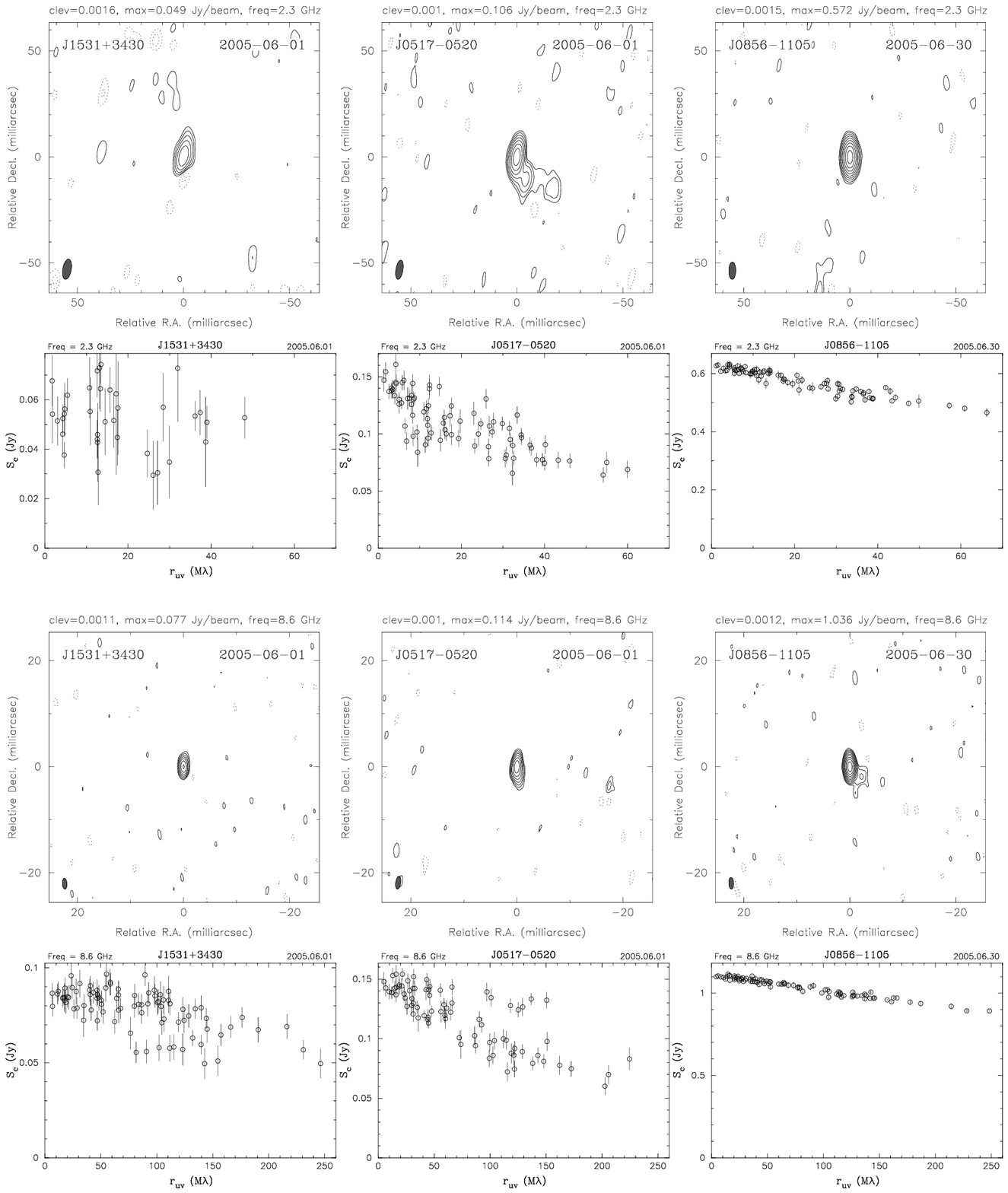}
                   }
        \ifpre \par \vspace{7ex} \par \hphantom{a} \fi
   \fi
   
   \figcaption{ \ifsub \footnotesize \fi \ifdraft \footnotesize \fi
        From top to bottom. Naturally weighted CLEAN images at S-band
        (2.3~GHz). The lowest contour levels (2 steps) on images are
        plotted at ``clev'' levels (Jy/beam), the peak brightness~--- ``max''
        values (Jy/beam). The dashed contours indicate negative flux.
        The beam is shown in the bottom left corner of the images.
        Dependence of the correlated flux density at S-band versus
        projected spacings. The error bars were computed on the basis
        of correlated flux density scatter around mean value. Naturally
        weighted CLEAN images at X-band (8.6~GHz). Dependence of the
        correlated flux density at X-band versus projected spacings.
        \label{f:images}
    }
\end{figure*}
\fi

\section{The VCS4 catalogue}
\label{s:catalogue}

\ifpre
  \begin{deluxetable*}{ c l l l r r r r r r r r r l}
\else
  \begin{deluxetable}{ c l l l r r r r r r r r r l}
\fi
\tablecaption{\rm The VCS4 catalogue \label{t:cat}}
\tabletypesize{\scriptsize}
\ifpre \relax \else \rotate \fi
\tablehead{
\vspace{2.0ex} \\
   \colhead{} &
   \multicolumn{2}{c}{Source name}                   &
   \multicolumn{2}{c}{J2000.0 Coordinates}           &
   \multicolumn{3}{c}{Errors (mas)}                  &
   \colhead{}                                        &
   \multicolumn{4}{c}{Correlated flux density (Jy)}  &
   \colhead{}
   \vspace{0.5ex} \\
   \multicolumn{9}{c}{}         &
   \multicolumn{2}{c}{8.6 GHz}  &
   \multicolumn{2}{c}{2.3 GHz}  \vspace{0.5ex} \\
   \vspace{0.5ex} \\
   \colhead{Class}    &
   \colhead{IVS}      &
   \colhead{IAU}      &
   \colhead{Right ascension} &
   \colhead{Declination}     &
   \colhead{$\Delta \alpha$} &
   \colhead{$\Delta \delta$} &
   \colhead{Corr}   &
   \colhead{\# Obs} &
   \colhead{Total } &
   \colhead{Unres } &
   \colhead{Total } &
   \colhead{Unres } &
   \colhead{Band}
   \vspace{0.5ex} \\
   \colhead{(1)}    &
   \colhead{(2)}    &
   \colhead{(3)}    &
   \colhead{(4)}    &
   \colhead{(5)}    &
   \colhead{(6)}    &
   \colhead{(7)}    &
   \colhead{(8)}    &
   \colhead{(9)}    &
   \colhead{(10)}   &
   \colhead{(11)}   &
   \colhead{(12)}   &
   \colhead{(13)}   &
   \colhead{(14)}
   }
\startdata
\vspace{2.0ex} \\
$-$ & 0006$-$363 & J0008$-$3601 & 00 08 33.661411 & $-$36 01 25.05213 &    6.20  & 32.64  &    0.697  &     8  &$  0.07 $ & \nodata & $  0.22 $ &  0.10   &  X/S \vspace{0.5ex} \\
C   & 0010$+$336 & J0012$+$3353 & 00 12 47.382197 & $+$33 53 38.47157 &    0.58  &  0.84  & $-$0.508  &    41  &$  0.16 $ & 0.13    & $  0.07 $ &  0.06   &  X/S \vspace{0.5ex} \\
C   & 0012$+$319 & J0015$+$3216 & 00 15 06.147414 & $+$32 16 13.30953 &    0.33  &  0.55  & $-$0.211  &    86  &$  0.20 $ & 0.11    & $  0.13 $ &  0.09   &  X/S \vspace{0.5ex} \\
C   & 0021$-$084 & J0024$-$0811 & 00 24 00.672734 & $-$08 11 10.04881 &    1.08  &  2.13  & $-$0.252  &    31  &$  0.10 $ & 0.08    & $  0.09 $ &  0.07   &  X/S \vspace{0.5ex} \\
C   & 0024$-$114 & J0026$-$1112 & 00 26 51.443027 & $-$11 12 52.42503 &    0.96  &  1.68  &    0.219  &    30  &$  0.11 $ & 0.05    & $  0.19 $ &  0.09   &  X/S \vspace{0.5ex} \\
C   & 0032$+$612 & J0035$+$6130 & 00 35 25.310617 & $+$61 30 30.76144 &    0.93  &  0.60  &    0.200  &    71  &$  0.14 $ & 0.06    & $  0.23 $ &  0.07   &  X/S \vspace{0.5ex} \\
$-$ & 0033$-$088 & J0035$-$0835 & 00 35 46.250383 & $-$08 35 54.04258 &    0.83  &  1.57  & $-$0.240  &    86  &$  0.13 $ & 0.12    & $  0.06 $ &  0.06   &  X   \vspace{0.5ex} \\
C   & 0052$-$201 & J0054$-$1953 & 00 54 32.948443 & $-$19 53 01.00202 &    0.42  &  0.83  & $-$0.091  &    59  &$  0.16 $ & 0.10    & $  0.20 $ &  0.05   &  X/S \vspace{0.5ex} \\
C   & 0052$-$125 & J0055$-$1217 & 00 55 11.782596 & $-$12 17 57.09709 &    0.25  &  0.46  & $-$0.220  &    90  &$  0.24 $ & 0.18    & $  0.21 $ &  0.15   &  X/S \vspace{0.5ex} \\
$-$ & 0057$+$101 & J0059$+$1022 & 00 59 46.769108 & $+$10 22 40.42531 &   16.93  & 17.91  &    0.797  &     2  & \nodata  & \nodata & \nodata   & \nodata &  S   \vspace{0.5ex} \\
\enddata
\tablecomments{ \rm
               Table~\ref{t:cat} is presented in its entirety in the electronic
               edition of the Astronomical Journal. A portion is shown here
               for guidance regarding its form and contents. Units of right
               ascension are hours, minutes and seconds, units of declination
               are degrees, minutes and seconds.}
\ifpre
  \end{deluxetable*}
\else
  \end{deluxetable}
\fi

  The VCS4 catalogue is listed in Table~\ref{t:cat}. The first column gives
source class: ``C'' if the source can be used as a calibrator, ``--'' if it
cannot. The second and third columns give IVS source name (B1950 notation),
and IAU name (J2000 notation). The fourth and fifth columns give source
coordinates at the J2000.0 epoch. Columns /6/ and /7/ give inflated source
position uncertainties in right ascension and declination (without
$\cos\delta$ factor) in mas, and column /8/ gives the correlation coefficient
between the errors in right ascension and declination. The number of group
delays used for position determination is listed in column /9/. Columns /10/
and /12/ give the estimate of the flux density integrated over entire map
in Jansky at X and S~band respectively. This estimate is computed as a sum 
of all CLEAN components if a CLEAN image was produced. If we did not have
enough detections of visibility function to produce a reliable image,
the integrated flux density is estimated as the median of the correlated 
flux density measured at projected spacings less than 25 and 7~M$\lambda$ 
for X~and S~bands respectively. The integrated flux density has a meaning 
of the source total flux density with spatial frequencies less than
4~M$\lambda$ at X~band and 1~M$\lambda$ at S~band filtered out, or by another 
words this is the flux density from all details of a source with size less 
than 50~mas at X~band and 200~mas at S~band. Column /11/ and /13/ give
the flux density of unresolved components estimated as the median of
correlated flux density values measured at projected spacings greater
than 170~M$\lambda$ for X~band and greater than 45~M$\lambda$ for
S~band. For some sources no estimates of the integrated and/or unresolved 
flux density are presented, because either no data were collected at the
baselines used in calculations, or these data were unreliable. Column /14/
gives the data type used for position estimation: X/S stands for
ionosphere-free linear combination of X and S wide-band group delays;
X stands for X~band only group delays; and S stands for S~band only group
delays. Some sources which yielded less than 8 pairs of X and S band group
delay observables had 2 or more observations at X and/or S band observations.
For these sources either X-band or S-band only estimates of coordinates are
listed in the VCS4 catalogue, whichever uncertainty is less.

\ifpre
\begin{figure*}[ht]
   \ifsub
         \epsscale{1.0}\plottwo{f8a.eps}{f8b.eps}
         \par\vspace{1ex}\par
      \else
         \centerline{
                       \epsfclipon
                       \epsfxsize=0.45\textwidth \epsffile{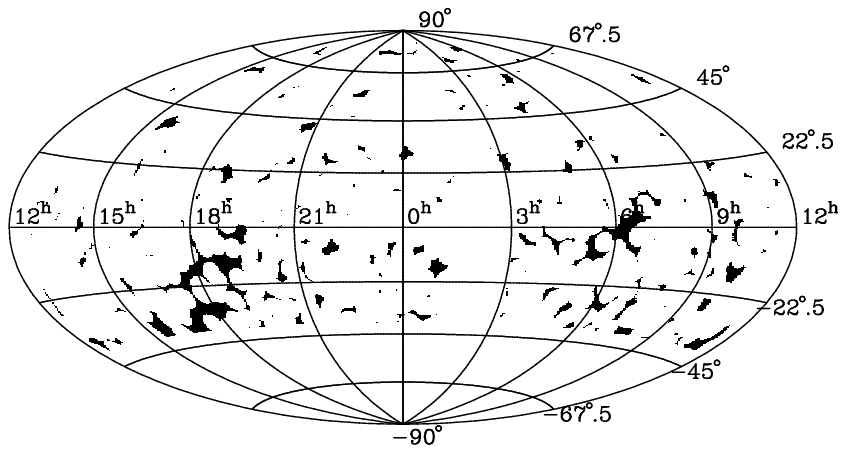}
                       \hspace{0.09\textwidth}
                       \epsfxsize=0.45\textwidth \epsffile{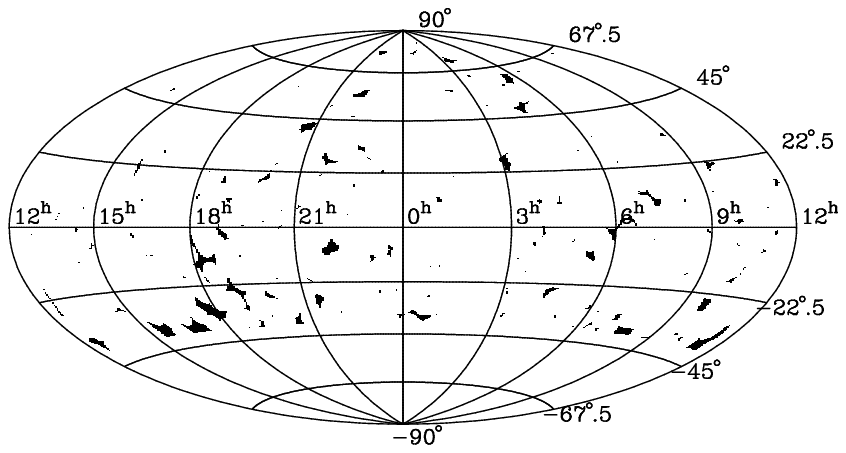}
                     }
   \fi

   \figcaption{ \ifsub \footnotesize \fi \ifdraft \footnotesize \fi
            Source sky density before (left) and after (right) the VCS4
            observing campaign. The area above $-40\degr\ $ declination
            with calibrator source density less than one source within
            a 4\degr\ radius disk is shown in black.
            \label{f:skymap}
           }
   \label{f:skydens}
\end{figure*}
\fi

  In addition to this table, the html version of this catalogue is posted
on the Web\footnote{\url{http://gemini.gsfc.nasa.gov/vcs4}}. For each source
it has 8 links: to a pair of postscript maps of the source at X and S-band;
to a pair of plots of correlated flux density as a function of the length
of the baseline projection to the source plane; to a pair of fits files
with CLEAN components of naturally weighted source images; and to a pair
of fits files with calibrated $uv$ data. The coordinates and the plots are
also accessible from the NRAO VLBA Calibration Search
web-page\footnote{\url{http://www.vlba.nrao.edu/astro/calib}}.

  Figure~\ref{f:images} presents examples of naturally weighted contour CLEAN
images as well as estimates of the correlated flux density versus projected
spacings. The S~band data for the source J1531+3430 is an example of one of
the weakest sources which we still were able to image. It turned out to be
an object with an inverted radio spectrum of the milliarcsecond emission.
The images of J0517-0520 show the compact structure with jet components which
were detected on the level of several mJy/beam only. The source J0856-1105
has the highest VLBI flux density at X~band among all the new VCS4 objects
and will be useful for geodetic applications along with a few tens of other
compact VCS4 objects with high flux density at VLBA baselines.

\ifpre \relax \else
\begin{figure*}[ht]
   \ifsub
        \epsscale{0.90}\plotone{f7.eps}
      \else
        \centerline{ \epsfclipon
                     \ifdraft \epsfxsize=0.95\textwidth \fi
                     \ifpre \epsfxsize=1.00\textwidth \fi
                     \epsffile{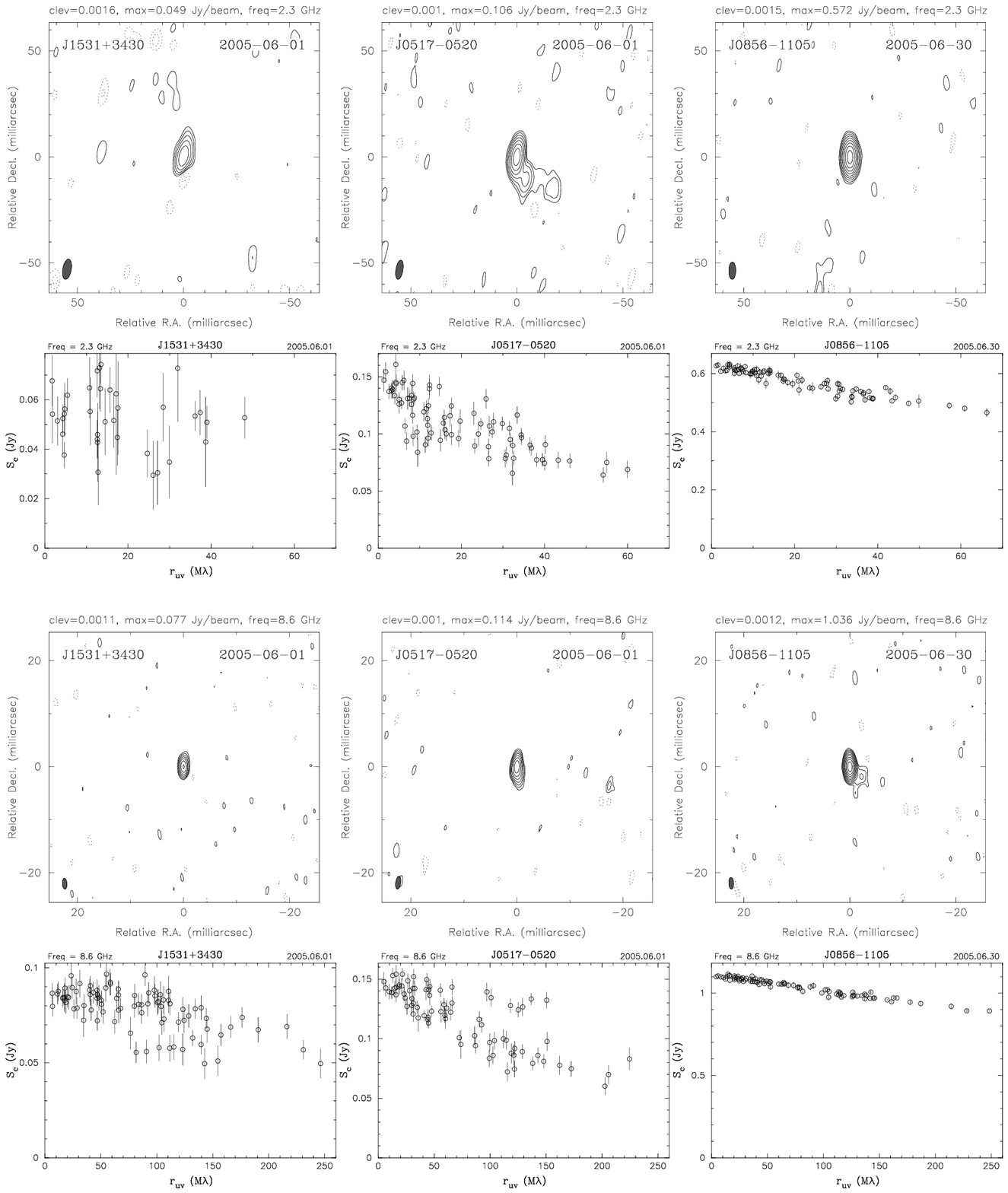}
                   }
   \fi

   \figcaption{ \ifsub \footnotesize \fi \ifdraft \footnotesize \fi
        From top to bottom. Naturally weighted CLEAN images at S-band
        (2.3~GHz). The lowest contour levels (2 steps) on images are
        plotted at ``clev'' levels (Jy/beam), the peak brightness~--- ``max''
        values (Jy/beam). The dashed contours indicate negative flux.
        The beam is shown in the bottom left corner of the images.
        Dependence of the correlated flux density at S-band versus
        projected spacings. The error bars were computed on the basis
        of correlated flux density scatter around mean value. Naturally
        weighted CLEAN images at X-band (8.6~GHz). Dependence of the
        correlated flux density at X-band versus projected spacings.
        \label{f:images}
    }
\end{figure*}
\fi

\section{Summary}
\label{s:summary}

   The VCS4 Survey has added 258 new sources not previously observed
with VLBI. Among of them, 199 sources turned out to be suitable as phase
referencing calibrators and as target sources for geodetic applications. The
area with source density less than one calibrator within a disk of radius
$\Deg{4}$ in any given direction on the sky with $ \delta > \Deg{-40} $ was
reduced from \mbox{4.6\,\%} to \mbox{1.9\,\%} (refer to
figure~\ref{f:skydens} and table~\ref{t:density}). After processing the VCS4
experiments, the total number of calibrators has grown from 2268 to 2472.
This pool of calibrators was formed from analysis of 18 VLBI Calibrator
Survey and 3976 twenty four hour IVS astrometry and geodesy observing
sessions.

\ifpre \relax \else
\begin{figure*}[ht]
   \ifsub
         \epsscale{1.0}\plottwo{f8a.eps}{f8b.eps}
      \else
         \centerline{
                       \epsfclipon
                       \epsfxsize=0.45\textwidth \epsffile{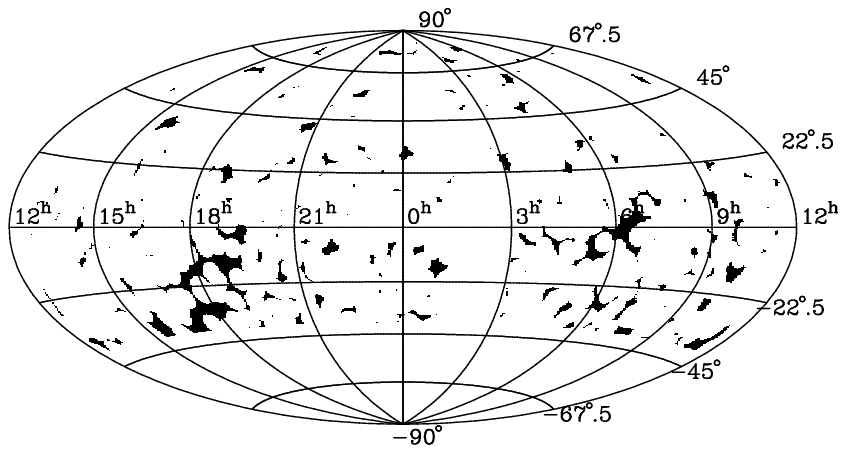}
                       \hspace{0.09\textwidth}
                       \epsfxsize=0.45\textwidth \epsffile{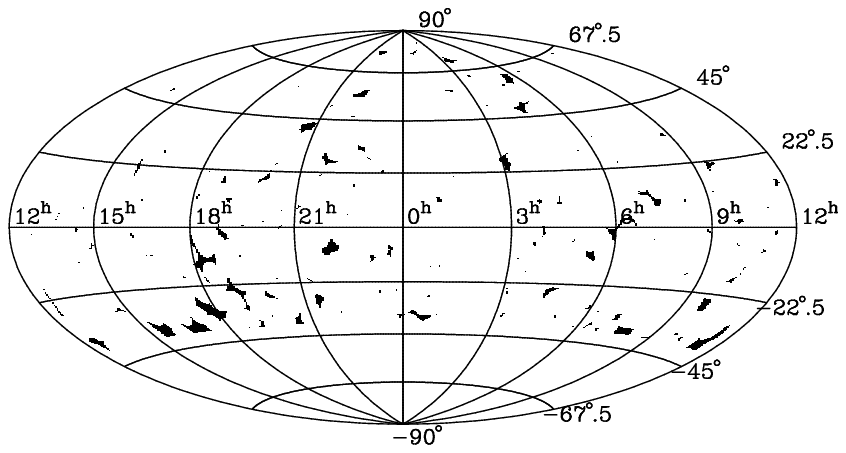}
                     }
   \fi

   \figcaption{ \ifsub \footnotesize \fi \ifdraft \footnotesize \fi
            Source sky density before (left) and after (right) the VCS4
            observing campaign. The area above $-40\degr\ $ declination
            with calibrator source density less than one source within
            a 4\degr\ radius disk is shown in black.
            \label{f:skymap}
           }
   \label{f:skydens}
\end{figure*}
\fi

   The strategy of source selection based on a~priori probabilities
of source detection developed in this study was successful.
The conservative a~priori estimate of detection rate was 57\%, while
the a~posteriori detection rate is 70\,\%. The a~priori estimate of the area
with the calibrator density less than one object within a disk of radius
$\Deg{4}$ was \mbox{1.8\,\%}, while the a~posteriori value is \mbox{1.9\,\%}.

\begin{table}[b]
  \caption{ \footnotesize\rm
           The probability to find a calibrator at any given direction within
           a circle of a given radius in selected declination zones.}
  \begin{center}
     \begin{tabular}{c  @{\qquad} r @{\qquad} r @{\qquad} r}
        \hline
            Search circle & \nnntab{c}{Declination zone} \\
            radius        & [$\Deg{-90}$, $\Deg{-40}]$  &
                            [$\Deg{-40}$, $\Deg{-30}]$ &
                            [$\Deg{-30}$, $\Deg{+90}]$ \\
         \hline
         $\Deg{4}\!.0$  &  63.6\,\%\Hem &  92.2\,\%\Hem  &  98.6\,\%\Hem  \\
         $\Deg{3}\!.5$  &  55.4\,\%\Hem &  86.0\,\%\Hem  &  96.0\,\%\Hem  \\
         $\Deg{3}\!.0$  &  44.1\,\%\Hem &  76.4\,\%\Hem  &  89.7\,\%\Hem  \\
         $\Deg{2}\!.5$  &  33.2\,\%\Hem &  56.6\,\%\Hem  &  77.9\,\%\Hem  \\
         $\Deg{2}\!.0$  &  22.7\,\%\Hem &  46.6\,\%\Hem  &  60.7\,\%\Hem  \\
         $\Deg{1}\!.5$  &  13.6\,\%\Hem &  29.4\,\%\Hem  &  40.0\,\%\Hem  \\
         $\Deg{1}\!.0$  &   6.3\,\%\Hem &  14.2\,\%\Hem  &  20.1\,\%\Hem  \\
         $\Deg{0}\!.5$  &   1.6\,\%\Hem &   3.7\,\%\Hem  &   5.4\,\%\Hem  \\
         \hline
      \end{tabular}
   \end{center}
\tablecomments{\rm
All sources from 3976 IVS astrometric and geodetic sessions and
18 VCS1, VCS2, VCS3 and VCS4 experiments with the VLBA which are
classified as calibrators are taken into account.
}
   \label{t:density}
\end{table}

   The sky calibrator density for different radii of a search circle and
for three different declination zones is presented in table~\ref{t:density}.
A further search for new calibrators to decrease the area with insufficient
sky calibrator density would be considerably less efficient, because
the majority of the flat spectrum sources in those areas have already been
observed. The remaining sources are the steep spectrum sources, or sources
with unknown spectral index, or very weak objects. According to
figure~\ref{f:prob5} the lack of information about spectral index of
potential candidates makes the detection rate of almost any sample about
the same.

  At the same time, some relatively bright flat spectrum sources remained in
the zones where there is at least one calibrator in a disk with radius
$\Deg{4}$. A sample of 675 such sources was observed in July 2005 with the
VLBA in a follow-up VCS5 experiment (Y.Y.~Kovalev et~al. in preparation).
The addition of these sources to the VLBA Calibrator list will not
affect the calibrator sky density with the search radius of $\Deg{4}$, but
will increase the density of calibrators with smaller search circles, which
will be beneficial for many applications, for instance, the VERA
project \citep{VERA}.

  New VCS4 detected sources were observed in 2005 with the Russian Academy
of Sciences 600~meter ring radio telescope RATAN for measurement of their
{\em instantaneous} 1--22 GHz spectra (Y.Y.~Kovalev et~al.\ in preparation).
These data will allow us to determine compactness of these sources and their
accurate spectral indexes. Using this information one can further improve
the estimate of the a~priori probability of source detection with VLBI
as a function of source flux and spectral index which is essential for
source selection in future surveys.

\acknowledgments

   The National Radio Astronomy Observatory is a facility of the
National Science Foundation operated under cooperative agreement by
Associated Universities, Inc. This work was done while L.~Petrov and
D.~Gordon worked for NVI, Inc. and Raytheon, respectively, under NASA
contract NAS5--01127. The authors are thankful to G.~Lipunova for valuable
comments. \mbox{RATAN--600} observations were partly supported by the
Russian State Program ``Astronomy'' and the Russian Ministry of Education
and Science, the NASA JURRISS Program (project W--19611), and
the Russian Foundation for Basic Research (projects 01--02--16812 and
05--02--17377). The authors made use of the database CATS \citep{cats}
of the Special Astrophysical Observatory. This research has made use of
the NASA/IPAC Extragalactic Database (NED) which is operated by the Jet
Propulsion Laboratory, California Institute of Technology, under
contract with the National Aeronautics and Space Administration.


\begin{thebibliography}{99}

   \bibitem[Beasley et~al.(2002)]{vcs1} Beasley,~A.~J., Gordon,~D.,
            Peck,~A.~B., Petrov,~L., MacMillan,~D.~S., Fomalont,~E.~B., \&
            Ma,~C. 2002, \apjs, 141, 13

   \bibitem[Fomalont et~al.(2003)]{vcs2}
            Fomalont,~E., Petrov,~L., McMillan,~D.~S., Gordon,~D., Ma,~C.
            2003, \aj, 126, 2562


   \bibitem[F\"urst et al.(1990)]{11cmb}
            F\"urst,~E., Reich,~W., Reich,~P., \& Reif,~K. 1990, \aaps, 85, 805

   \bibitem[Greisen (1988)]{aips}
            Greisen,~E.~W., 1988, in Acquisition,
            Processing and Archiving of Astronomical Images, ed. G.~Longo \&
            G.~Sedmak (Napoli: Osservatorio Astronomico di Capodimonte), 125

   \bibitem[Honma et al.(2003)]{VERA}
        Honma,~M., Fujii,~T., Hirota,~T., Horiai,~K.,
	Iwadate,~K., Jike,~T., Kameya,~O., Kamohara,~R.,
	Kan-Ya,~Y., Kawaguchi,~N., Kobayashi,~H., Kuji,~S.,
	Kurayama,~T., Manabe,~S., Miyaji,~T., Nakashima,~K.,
	Omodaka,~T., Oyama,~T., Sakai,~S., Sakakibara,~S.-I.,
	Sato,~K., Sasao,~T., Shibata,~K.~M.,~Shimizu,~R.,
	Suda,~H., Tamura,~Y., Ujihara,~H., \& Yoshimura,~A.
        2003, \pasj, 55, L57

   \bibitem[Jackson et~al.(2002)]{parkes-qJy}
            Jackson,~C.~A., Wall,~J.~V., Shaver,~P.~A., Kellermann,~K.~I.,
            Hook,~I.~M., \& Hawkins,~M.~R.~S. 2002, \aap, 386, 97

   \bibitem[Kellermann, Pauliny-Toth, \& Davis(1968)]{kellermann}
          Kellermann,~K.~I., Pauliny-Toth,~I.~I.~K., \& Davis,~M.~M.
          1968, Astrophysical Letters, 2, 105

   \bibitem[Kellermann \& Owen(1988)]{ko88}
          Kellermann,~K.~I. \& Owen,~F.~N. 1988, Galactic and Extragalactic
          Radio Astronomy, ed.~by G.~L.~Verschur \& K.~I.~Kellermann
          (Springer Verlag), 563

   \bibitem[Kellermann et~al.(1998)]{2cm}
       Kellermann,~K.~I., Vermeulen,~R.~C., Zensus~J.~A., \& Cohen~M.~H.
          1998, \aj, 115, 1295

   \bibitem[Kovalev et~al.(1999)]{Kovalev_etal99}
            Kovalev,~Y.~Y., Nizhelsky,~N.~A., Kovalev,~Yu.~A., Berlin,~A.~B.,
            Zhekanis,~G.~V., Mingaliev,~M.~G., \& Bogdantsov,~A.~V.
            1999, \aaps, 139, 545

   \bibitem[Kovalev et~al.(2005)]{Kovalev_etal2005}
            Kovalev, Y.~Y., Kellermann, K.~I., Lister, M.~L.,
            Homan, D.~C., Vermeulen, R.~C., Cohen, M.~H., Ros, E.,
            Kadler, M., Lobanov, A.~P., Zensus, J.~A.,
            Kardashev, N.~S., Gurvits, L.~I., Aller, M.~F., \&  Aller, H.~D.\
            2005, \aj, 130, in press (astro-ph/0505536)

   \bibitem[Ma et~al.(1998)]{icrf98}
	Ma,~C., Arias,~E.~F., Eubanks,~T.~M., Fey,~A.~L., Gontier,~A.-M.,
        Jacobs,~C.~S., Sovers,~O.~J., Archinal,~B.~A., \& Charlot,~P.,
        1998, \aj, 116, 516

   \bibitem[Fey et~al.(2004)]{icrf-ext2-2004} Fey,~A.L., Ma,~C.,
            Arias,~E.~F., Charlot,~P., Feissel-Vernier,~M., Gontier,~A.-M.,
            Jacobs,~C.~S., Li,~J., MacMillan,~D.~S. 2004, \aj, 127, 3587.

   \bibitem[Matveenko et~al.(1965)]{mat65}
            Matveenko, L.~I., Kardashev, N.~S., \& Sholomitskii, G.~B.\
            1965, Izvestia VUZov.\ Radiofizika, 8, 651
            (English transl.\ Soviet Radiophys., 8, 461)

   \bibitem[Mingaliev et~al.(2001)]{RATAN_NCP}
           Mingaliev, M.~G., Stolyarov, V.~A., Davies, R.~D.,
	   Melhuish, S.~J., Bursov, N.~A., \&  Zhekanis, G.~V.\
	   2001, \aap, 370, 78

   \bibitem[Pearson et~al.(1994)]{difmap-script}
            Pearson,~T.~J., Shepherd,~M.~C., Taylor,~G.~B., \& Myers, S.~T.
            1994, \baas, 185, 0808

   \bibitem[Petrov et~al.(2005)]{vcs3}
     Petrov,~L., Kovalev,~Y.Y., Fomalont,~E., Gordon,~D. 2005, \aj, 129, 1163.

   \bibitem[Slysh(1963)]{slysh}
   Slysh,~V.I. 1963, Nature, 199, 682

   \bibitem[Verkhodanov et~al.(1997)]{cats}
            Verkhodanov,~O.~V., Trushkin,~S.~A., Andernach,~H., \&
            Chernenkov,~V.~N. 1997, in ASP Conf. Ser. 125,
            Astronomical Data Analysis Software and Systems VI,
            ed.~by G.~Hunt, \& H.~E.~Payne (San Francisco: ASP), 322

  \bibitem[Shepherd(1997)]{difmap}
          Shepherd,~M.~C. 1997, in ASP Conf.~Series. 125,
          Astronomical Data Analysis Software and Systems~VI,
          ed.~by G.~Hunt \& H.~E.~Payne (San Francisco: ASP), 77

   \bibitem[Winn, Patnaik, \& Wrobel(2003)]{phcal-south}
	    Winn,~J.~N., Patnaik,~A.~R., \& Wrobel,~J.~M. 2003, \apjs, 145, 83

   \bibitem[Wright \& Otrupcek(1990)]{pkscat90a}
            Wright, A.~E., \& Otrupcek, R.\ 1990,
            PKSCAT90 Radio Source Catalogue and Sky Atlas
            (Epping: Australia Telesc. Natl. Facility)

   \bibitem[Wrobel \& Ulvestad (2005)]{VLBA_summ}
         Wrobel,~J.~M. \& Ulvestad,~J.~S. 2004, VLBA status summary, NRAO

\end{thebibliography}
\end{document}